\documentclass[draft,showpacs,preprintnumbers,amsmath,amssymb,superscriptaddress]{revtex4-1}

\newcommand{\be}{\begin{equation}}
\newcommand{\ee}{\end{equation}}

\usepackage{multirow}
\usepackage{amsthm}
\newtheorem{mytheorem}{Theorem}
\begin{document}
\title{On Stationary Axially Symmetric  Solutions in Brans-Dicke Theory}

\begin{abstract}

Stationary, axially symmetric Brans-Dicke-Maxwell  solutions are reexamined in the framework of the Brans-Dicke (BD) theory. We see that, employing a particular parametrization of the standard axially symmetric metric simplifies the procedure of obtaining the Ernst equations for axially symmetric electrovacuum space-times for this theory.  This analysis also  permits us to construct a two parameter extension  in both Jordan and Einstein frames of an old solution generating technique frequently used to construct axially symmetric solutions for BD theory from a seed solution of general relativity. As applications of this technique, several known and new solutions are constructed including a general axially symmetric BD-Maxwell solution of Plebanski-Demianski with vanishing cosmological constant, i.e. the Kinnersley solution and general magnetized Kerr-Newman--type solutions. Some physical properties and the circular motion of  test particles  for a particular subclass of Kinnersley solution, i.e., a Kerr-Newman-NUT--type solution for BD theory, are also investigated in some detail. 

\end{abstract}


\author{P\i nar Kirezli}
\email{pkirezli@nku.edu.tr}
\affiliation{Department of Physics, Faculty of Arts and Sciences, Nam\i k Kemal University,  Tekirda\u g 59030,  Turkey}
\affiliation{Department of Physics, Faculty of Arts and Sciences, Marmara University,  Istanbul 34722, Turkey}
\author{\"Ozg\"ur Delice}
\email{ozgur.delice@marmara.edu.tr}
\affiliation{Department of Physics, Faculty of Arts and Sciences, Marmara University,  Istanbul 34722, Turkey}
\date{\today}

\maketitle

\section{Introduction}

Brans-Dicke (BD) scalar-tensor theory \cite{BD} is the most studied alternative theory of gravity generalizing general relativity (GR)  in a consistent way by introducing a scalar  field replacing Newton's gravitational constant. Being one of the most straightforward extension of GR,  this theory draws a lot of interest and  helps to test various aspects of GR. The peculiar differences were noted in the exact solutions of this theory compared to the similar solutions of GR. For example, static spherically symmetric vacuum solutions of BD theory \cite{Brans}, unlike GR, do not describe asymptotically flat black holes \cite{Hawking} resulting from gravitational collapse, unless the scalar field becomes a constant. This is due to the fact that corresponding solutions in BD theory cannot meet three important criteria, namely, asymptotic flatness, regularity at the horizon and the weak energy condition, simultaneously. After the discovery of spherically and axially symmetric exact solutions, it was realized that there are ranges of parameters in which these solutions describe black holes \cite{Campanelli,Bronnikov,Bhadra,KimBH,Park} where the first two criteria are met but the third one is not. However, these ranges are at the nonphysical negative values of the BD parameter, $\omega$, where the kinetic term becomes negative.  Hence,  it is concluded that the black holes of BD theory obeying these three conditions  must be the same as the black holes of GR, requiring a  constant BD scalar. Although they are the same, their perturbations can behave differently \cite{Sotiriu} owing to the fact that they are the solutions of different theories. Thus, the differences between perturbations of  GR and BD theories for the same black hole solution might be another tool \cite{Sotiriu,Cardoso} to test BD theory against GR.

Obtaining exact solutions of any theory is important for several reasons, such as, for comparisons with observational results or for obtaining the outcomes of the theory under consideration. One important class of these solutions with a great physical importance are the family of stationary, axially symmetric solutions,  
since the  gravitational field of compact celestial objects such as stars, galaxies and  black holes can be represented by such solutions. Investigating their properties may have some important astrophysical effects  such as  the existence and stability of orbits of planets and stars around such objects. In order to discuss these important properties of a gravitation theory,  exact or approximate solutions must be obtained. 
 For the BD theory  we are considering, the field equations are more complex then those of GR, due to  the presence of the extra scalar field.  Even in  GR, several solution generating techniques are developed  due to the complexity of the field equations. Thus, in order to obtain exact solutions of BD theory, similar techniques were used. One of the methods frequently employed is making use of the Ernst equations \cite{Ernst1,Ernst2} derived from BD field equations  \cite{chatterjee,tsuchida,Astorino1}. Another method considers conformal transformation properties of scalar-tensor theories and using the equivalence of vacuum or Maxwell BD theory with Einstein-Maxwell scalar solutions \cite{Buchdahl1,Janis}.

 Besides these methods, following the earlier works on this subject \cite{Buchdahl,Mcintosh,Tupper,Tiwari,ErisGurses},  an identification technique  is presented by  Nayak and Tiwari \cite{Nayak} for axially symmetric vacuum solutions and generalized  for Maxwell vacuum fields by Sing and Rai \cite{Singh}. The main idea of this method is to employ a standard stationary, axially symmetric spacetime metric and the field variables sharing the symmetries of this metric and  compare the field equations of both theories to find some transformations 
  to reduce the field equations of BD theory to those of GR. Once the transformations linking general axially symmetric solutions of GR theory to BD theory are found, one can effortlessly obtain  new BD solutions  for a given axially symmetric solution of GR. Thus, this method, which we call the Tiwari-Nayak-Singh-Rai (TNSR) method, is the most direct one. Using this method, Kerr-Newman \cite{KimBH} and black dihole type \cite{KimLee} solutions were constructed.  On the other hand,  one caveat for the present form of the TNSR  method is that  it contains only one extra parameter, the BD parameter, $\omega$, in addition to the parameters of the original GR solution. However, various vacuum or Maxwell vacuum solutions such as the spherically symmetric BD vacuum solutions \cite{Brans}, the Kerr-like solution presented in \cite{Mcintosh}, and cylindrically symmetric static vacuum  or Einstein-Maxwell solutions \cite{Baykal} all  contain an extra arbitrary parameter in addition to $\omega$.  Moreover, a recent work  \cite{DeliceAkyar} presenting  cylindrically symmetric Einstein-Rosen--type vacuum solutions also employed a similar technique involving two parameters. 
 This fact implies that the TNSR method is not in its most general form.

 In this paper, our main aim is to investigate general stationary axially symmetric solutions of BD-Maxwell theory by first obtaining a more general extension of  TNSR method with one extra parameter that can be regarded as a  measure of departure from GR. 
 In order to do this, we first analyze the field equations for the  general stationary axially symmetric spacetime given in the cylindrical coordinates in Sec. II. By employing a nonstandard reparametrization for the metric functions of a standard axially symmetric metric, we are able to obtain the Ernst equations from the field equations more easily. Analyzing this equation and remaining field equations together  permit us to construct a two parameter extended version  of the TNSR method in Sec. III, in both Jordan and Einstein frames.  
 Using this extended  method we will first  give some simple examples to show how this method works. Then  we will  obtain a BD version of  the  general axially symmetric Einstein-Maxwell type $D$ solution, known as the Plebanski-Demianski solution in the case of the vanishing cosmological constant \cite{Plebansky,Debever,Kinnersley} in Sec. IV. We will also investigate some  physical properties and the geodesics  of a particular subclass of these solutions, namely, the BD  version of a Kerr-Newman-NUT type solution, where the detailed calculations of the geodesic motion will be given in Appendix \ref{circgeo}.  Sec. V is devoted to the discussion  of ways of obtaining magnetized solutions for BD theory from a seed solution. Thanks to the technical convenience provided by   this method,  we have obtained the general magnetized BD Kerr-Newman--type solution with a little effort using the corresponding GR solution and the extended TNSR method. As a last section in the paper, we  will discuss the  GR limit of these solutions in detail. This paper ends with a brief discussion.

\section{Field Equations and some solution generating techniques}

\subsection{Spacetime and field equations}

In this paper we consider Brans-Dicke scalar-tensor theory with a Maxwell field, described by the action in the Jordan frame as 
\begin{equation}
S_{JBD}=\int d^4 x \sqrt{-g} \left(\phi R-\frac{\omega}{\phi}g^{\mu\nu}\partial_\mu \phi \partial_\nu\phi -F_{\mu\nu}F^{\mu\nu}\right).
\end{equation}
The field equations of this action can be expressed as
\begin{eqnarray}
&& G_{\mu\nu}=\frac{T_{\mu\nu}}{\phi}+\frac{\omega}{\phi^2}\left(\phi_{,\mu} \phi_{,\nu}-\frac{1}{2}\phi^{,\lambda} \phi_{,\lambda} \right) 
-\frac{1}{\phi}\left(\phi_{;\mu;\nu}-g_{\mu\nu}\Box \phi \right),\label{BDFE}\\
&&\left(2\omega+3 \right)\,\Box \phi=T=0.\label{BDSE}
\end{eqnarray}
Here $ T_{\mu\nu}$  is the energy-momentum tensor of the Maxwell field given by
\begin{equation}
T_{\mu\nu}=2 \left(F_{\mu}^{\phantom{a}\alpha}F_{\nu\alpha}-\frac{1}{4} g_{\mu\nu}F^{\alpha\beta}F_{\alpha\beta}  \right),
\end{equation}
 and $T=T^{\mu}_{\phantom{a}\mu}$ is its trace which is vanishing in four dimensions. 

 A general stationary axially symmetric spacetime  can be represented with a metric in cylindrical coordinates in the canonical form as 
\begin{equation}\label{metricansatze}
ds^2=-e^{2U}(dt+\mathcal{A}d\varphi)^2+e^{2(K-U)}(d\rho^2+dz^2)+e^{-2U}W^2d\varphi^2,
\end{equation}
where the metric functions $U,K,\mathcal{A}$ and $W$ are the functions of $\rho$ and $z$. This metric admits two Killing vectors $\partial_t$ and $\partial_\varphi $. This metric can  also be expressed in terms of other relevant coordinate systems such as Boyer-Lindquist coordinates by some suitable coordinate transformations. We also consider a Maxwell field sharing the symmetries of spacetime  given by the following potential one-form:
\begin{equation}\label{Maxwelloneform}
A=A_0(\rho,z)\,dt+A_3(\rho,z)\, d\phi.
\end{equation}

 \subsection{Ernst equation for Brans-Dicke-Maxwell theory} 
 
Now we discuss how to reduce BD field equations into the Ernst equation for axially symmetric spacetimes. In order to do this, following  \cite{Charmousis,Astorino}, it is appropriate to use a different but equivalent metric Ansatz given by
 \begin{equation}\label{metricastorino}
ds^2=-\alpha e^{\frac{\Omega}{2}}(dt+\mathcal{A} d\varphi)^2+\alpha e^{-\frac{\Omega}{2}}d\varphi^2+\frac{e^{2\nu}}{\sqrt{\alpha}}(d\rho^2+dz^2), 
 \end{equation}
 which simplifies the forthcoming equations considerably. The relation between the metric functions of (\ref{metricansatze}) and (\ref{metricastorino}) are given by
\begin{equation}\label{relationsmetrics}
U=\frac{\Omega}{4}+\frac{1}{2}\ln\alpha,\quad K=\frac{\Omega}{4}+\nu+\frac{1}{4}\ln \alpha,\quad W=\alpha.
\end{equation}
Note that although $W$  and $\alpha$ are the same, we will keep both symbols and use whichever one is relevant, in order to remind ourselves which metric \emph{Ansatze} we are using.

In the presence of a Maxwell field of the form  given in (\ref{Maxwelloneform}), BD-Maxwell field equations can be written in a compact form  as 
\begin{eqnarray}
&&\nabla^2(\alpha\phi)=0,\label{EE1}\\
&&\frac{1}{2}\vec{\nabla}.\left(\alpha\phi e^{\Omega}\vec \nabla{\mathcal{A}}\right) + 2 e^{\Omega/2}\big[\mathcal{A} \left(\vec{\nabla} A_0\right)^2
-\vec{\nabla} A_0 \vec{\nabla} A_3  \big]=0,\label{EE2}\\
&&\frac{1}{2}\vec{\nabla}.\left(\alpha \phi \vec{\nabla}\Omega \right)
+\mathcal{A}\vec{\nabla}.\left( \alpha\phi e^{\Omega}\vec{\nabla}\mathcal{A} \right)
+ \alpha \phi e^{\Omega}(\vec{\nabla}\mathcal{A})^2
\nonumber \\&& \
+\, 2 e^{\frac{\Omega}{2}}\left[\mathcal{A}^2\left(\vec{\nabla}A_0\right)^2-\left(\vec{\nabla} A_3 \right)^2 \right]
-2 e^{-\frac{\Omega}{2}}\left(\vec \nabla A_0\right)^2 =0,~~~~~ \label{EE3}\\
&&\frac{1}{2} \alpha\phi e^{\Omega} \left(\vec{\nabla}\mathcal{A} \right)^2-\frac{1}{8}\alpha \phi \left(\vec{\nabla}\Omega\right)^2+\frac{1}{2}\nabla^2(\alpha\phi)-2\alpha\phi\nabla^2\nu 
-\frac{3}{2}\alpha
\nabla^2\phi-\omega\alpha\frac{(\vec{\nabla}\phi)^2}{\phi}=0,
\label{EE4}\\
&&\frac{e^{-2\nu}}{\sqrt{\alpha}}(2\omega+3)\vec{\nabla}\left(\alpha \vec{\nabla}\phi \right)=0,\label{BDE}\\
&&\vec\nabla.\left[e^{-\Omega/2}\vec{\nabla}A_0+\mathcal{A}e^{\Omega/2}\left(\vec{\nabla}A_3-\mathcal{A}\vec{\nabla}A_0 \right)\right]=0,\label{ME1}\\
&&\vec{\nabla}.\left[e^{\Omega/2}\left(\vec{\nabla}A_3-\mathcal{A}\vec{\nabla}A_0 \right) \right]=0. \label{ME2}
\end{eqnarray}
The above equations (\ref{EE1})-(\ref{EE4}) are obtained from  BD field equations (\ref{BDFE}), Eq.  (\ref{BDE}) is  the scalar field equation (\ref{BDSE})  and  (\ref{ME1}) and (\ref{ME2}) are 
 the  nontrivial Maxwell equations which  do not involve the scalar field $\phi$. Here the differential operators are the operators in the flat Minkowski spacetime, i.e., $\nabla=\partial_\rho \vec e^\rho +\partial_z \vec e^z+\partial_\varphi  \vec e^ \varphi$ such that $\nabla^2=\partial_\rho^2+\partial_z^2+\partial_\varphi^2$. 
Note that in Eqs. (\ref{EE1})-(\ref{EE3}) the functions $\alpha$ and $\phi$ are present only as products $\alpha\phi$.
 Following \cite{Ernst1,Ernst2,Astorino}  we can put   Eqs. (\ref{EE2}),(\ref{EE3}),(\ref{ME1}) and (\ref{ME2}) into two  complex equations given by
 \begin{subequations}
	 \begin{align}
	(\mbox{Re}~\varepsilon+|\Phi|^2)\frac{1}{\bar{\alpha}}\vec{\nabla}.(\bar{\alpha}\vec{\nabla}\varepsilon)=(\vec{\nabla}\varepsilon+2\Phi^*\vec{\nabla}\Phi).\vec{\nabla}\varepsilon, \label{eps1}\\
	(\mbox{Re}~\varepsilon+|\Phi|^2)\frac{1}{\bar{\alpha}}\vec{\nabla}.(\bar{\alpha}\vec{\nabla}\Phi)=(\vec{\nabla}\varepsilon+2\Phi^*\vec{\nabla}\Phi).\vec{\nabla}\Phi, \label{eps2}
	\end{align}
	\end{subequations}
 where the new function $\bar{\alpha}$ is given by 
 \begin{equation}
 \bar{\alpha}=\phi\,\alpha=\phi\, W,
 \end{equation}
 satisfying Eq. (\ref{EE1}), i. e., $\nabla^2\bar{\alpha}=0.$
  The complex potential $\Phi$ and the complex function 
 $\varepsilon$  are defined as
 \begin{equation}
	\varepsilon=f-|\Phi|^2+ih,
	\end{equation}
	\begin{eqnarray}
	\Phi=A_0+i\tilde{A}_3.
	\end{eqnarray}
	Here the scalar function $f$ is given by 
	\begin{eqnarray}
	f=e^{\Omega/2}\alpha\phi=e^{\Omega/2}\bar{\alpha},
	\end{eqnarray}
 and new vector potentials $\tilde A_3$ and $h$ are defined by
 \begin{subequations}
 \begin{align}\label{defA3tilde}
\vec e_\varphi\times \vec\nabla \tilde A_3=e^{\Omega/2}(\vec \nabla A_3-\mathcal{A} \nabla A_0),\\
\vec{e}_{\varphi}\times\vec{\nabla}h=e^{\Omega}\bar{\alpha} \vec{\nabla}\mathcal{A}-2\vec{e}_{\varphi}\times
	\mbox{Im}(\Phi^*\vec{\nabla}\Phi),
\end{align} 
\end{subequations}
  where $\vec e^\varphi$ belongs to the set of  basis vectors of  three-dimensional flat space $\left\{\vec e^\rho, \vec e^\varphi, \vec e^z \right\}$.

Equations (\ref{eps1}) and (\ref{eps2}) are actually in the same form as the corresponding equations  in GR \cite{Stephani}.
The real and imaginary parts of the equation (\ref{eps1}) are equivalent to  the equations (\ref{EE2}) and (\ref{EE3}) whereas the real and imaginary parts of  Eq. (\ref{eps2}) are equivalent to  Eqs. (\ref{ME1}) and (\ref{ME2}).

The field equation (\ref{EE1}), namely, $\nabla^2(\alpha\phi)=\nabla^2{\bar\alpha}=0$, permits us to employ the canonical  value  for $\bar\alpha$, such that we can set $\bar{\alpha}=\rho$. This brings  Eqs (\ref{eps1}) and (\ref{eps2}) into
\begin{subequations}
\begin{align}
	 \left(\varepsilon+\varepsilon^*+|\Phi|^2\right)\tilde \nabla^2\varepsilon=2\left(\vec{\nabla}\varepsilon+2\Phi^*\vec{\nabla}\Phi\right).\vec{\nabla}\varepsilon, \label{eps3}\\
	\left(\varepsilon+\varepsilon^*+|\Phi|^2\right)\tilde\nabla^2\Phi=2\left(\vec{\nabla}\varepsilon+2\Phi^*\vec{\nabla}\Phi \right).\vec{\nabla}\Phi, \label{eps4}
\end{align}
\end{subequations}
where now $\tilde \nabla^2$ is the Laplacian operator in the three-dimensional flat cylindrical coordinates. 
These equations are exactly  in the same form with the Ernst equation \cite{Ernst1,Ernst2} for GR in the presence of a Maxwell field. Thus, the choice of the metric form (\ref{metricastorino}) simplifies the calculations in obtaining Ernst equations for BD theory considerably.  
Actually, the above Ernst equations  for BD theory were already derived  before \cite{chatterjee,tsuchida,Astorino1} in the Einstein frame, and some exact solutions were obtained by integrating these equations.
Note that the above Ernst equations of BD theory in the Jordan frame can also be put in the the other useful  forms of the Ernst equations known in the literature with exactly the same transformations. However, we will not pursue this path further in this paper.  
Note also that, for the following discussion, we will retain the $\bar{\alpha}$ term  in order to cover some exact solutions where $\bar{\alpha}\neq \rho$.

 For  given Ernst potentials solving these two equations, the remaining field equations (\ref{EE1}),(\ref{EE4}) and (\ref{BDE}) must be integrated to obtain the full solution. The fact that the form of the BD Ernst equations given above are \emph{exactly} the same equations with the Ernst equations of GR  enables us to make an important  observation that any solutions of the Ernst equations of GR  are  also the solutions of BD theory formulated in the manner presented above. This requires the identifications
 \begin{equation} \bar{\alpha}=\alpha\phi=\alpha_E,\quad \Omega=\Omega_E,\quad A=A_E,\quad \mathcal{A}=\mathcal{A}_E,
\end{equation} 
where subscript $E$ represents any solution of the Ernst equation in GR.
  The above relations are due to the  unconventional parametrization (\ref{relationsmetrics}) of the stationary axially symmetric metric (\ref{metricastorino}).  
     Note that there is an arbitrariness in the metric function $\alpha$ and the scalar field $\phi$ for the choice $\bar{\alpha}=\alpha\phi$,  since  any choice of $\alpha$ and $\phi$  satisfying  Eq. (\ref{EE1}) is not guaranteed  to solve the equation (\ref{BDE}). Hence, among all the possible choices of $\alpha$ and  $\phi$ satisfying (\ref{EE1}), only the ones satisfying Eq. (\ref{BDE}) are the solutions of the BD field equations. Thus, although it is a powerful method, obtaining solutions from Ernst equations still requires some effort to solve the field equation (\ref{BDE}) and integrate  Eq. (\ref{EE4}) to  find $\nu$. Hereafter, we will employ a more direct method, yielding the solutions readily from known GR solutions, up to  possible coordinate transformation. Actually, as we have discussed in the Introduction, there is such a method in the literature, given in \cite{Nayak,Singh}, which successfully generated many BD(-Maxwell) solutions from the known solutions of GR theory. In the next subsection, a more general form of this method, containing one extra parameter apart from the BD parameter $\omega$, will be presented, with the help of the Ernst equations obtained in the above analysis. 

\section{Extended Tiwari-Nayak-Rai-Singh Method}

 It is usually difficult to directly solve the field equations of Einstein's or Brans-Dicke  theory of gravity to obtain exact solutions. Due to this difficulty, some solution generating techniques are developed. Here we  extend one of them, which enables one to generate an axially symmetric stationary or static solution of Brans-Dicke(-Maxwell) theory from a known solution of Einstein(-Maxwell) gravity  with the same symmetry  properties. The essence of the method is the following. One starts with general BD(-Maxwell) field equations of a metric  of the form (\ref{metricansatze}), where the scalar field $\phi$ and the Maxwell field share the symmetries of the metric,   with  metric and field variables  $K,U, \mathcal{A},W,A,\phi$. Then by investigating the transformations on these metric functions such that the field equations  exactly reduce to field equations of Einstein(-Maxwell) theory for the metric and field variables of GR with $K_E,U_E, \mathcal{A}_E,W_E,A_E$.

An immediate observation of the field equations (\ref{EE1})-(\ref{ME2}), as we have discussed above, is that in Eqs. (\ref{EE2}) and (\ref{EE3}) or equivalently in (\ref{eps1}) and (\ref{eps2}) the metric function $W(=\alpha)$ and the scalar field $\phi$  enters the expressions only as  products $\bar{\alpha}=\alpha \phi$, owing to the form of the metric \emph{Ansatz} (\ref{metricastorino}). Since, in the above choices we have considered, the form of the Ernst equations are similar for GR and BD theories, any solutions of this equation for GR can be considered as solutions of this equation in BD theory provided that the product $\alpha \phi$ is equal to the  corresponding metric function for Einstein-Maxwell theory $\alpha_E$, together with the remaining metric functions and Maxwell fields except $\nu$. Note that for this choice $\alpha\phi=\alpha_E$,  it is easy to show that the requirement that $\alpha$ and $\phi$  solve the remaining field equations  
(\ref{EE1}) and (\ref{BDE}) is met by the  \emph{Ansatze} $\alpha=\alpha_E^k$, $\phi=\alpha_E^{1-k}$ with $k$ being a constant parameter, measuring the departure from GR. Actually, the choice is well motivated, since this  \emph{Ansatz} is a general solution for the canonical value $\alpha_E=\rho$ with the choice $\phi=f(\rho)$, $\alpha=\rho/f(\rho)$. For these choices  Eq. (\ref{BDE}) becomes
\begin{equation}
\left(\frac{\rho f'}{f}\right)'=0,
\end{equation}
whose general solution is
\begin{equation}
f(\rho)=\rho^{c_0},
\end{equation}
and we have chosen the integration constant $c_0=1-k$ similar to some previous solutions \cite{DeliceAkyar,Baykal} exist in the literature. We will not consider the possibility of more general choices of $\alpha$ and $\phi$ in this paper.

 By a careful analysis we have obtained the following theorem.
\begin{mytheorem}
From any solution of the Einstein or the Einstein-Maxwell theory with the metric of the form (\ref{metricastorino}) labelled by a subscript $E$ and a possible Maxwell field for the Einstein-Maxwell case, i. e., $A_E=A_0(\rho,z) dt+A_3(\rho,z)d\phi$, the corresponding BD or BD-Maxwell solution can be obtained by the following transformations: 
\begin{subequations}
 \begin{align}
&\alpha=\alpha_E^{k},\quad \phi=\alpha_E^{1-k},\quad
\Omega=\Omega_E,\label{trans2} \\
& \mathcal{A}=\mathcal{A}_E, \quad A=A_E,\label{trans1}\\
& \nu=\nu_E+\frac{2\omega-(2\omega +3)k}{4}\ln\phi.\label{nutrans}
\end{align} 
\end{subequations} 
\end{mytheorem}
To prove this theorem we only need to prove the relation (\ref{nutrans}),  which can be easily obtained by applying the transformations (\ref{trans2}) and (\ref{trans1}) to BD-Maxwell field equations above  and solving for $\nu$.     
Thus, for this form  (\ref{metricastorino}) of the axially symmetric stationary spacetime, the transformation equations relating GR solutions to BD ones turn out to be  quite simple. The advantage of this form of the metric is that the Ernst equations of BD and GR theories  are  obtained straightforwardly  in these coordinates and the related developments for the Ernst equation are easily applicable to BD theory in this parametrization. However, the form of the metric (\ref{metricastorino}) is rather unfamiliar. Hence, in order to compare our results with the existing solutions in the  literature and also to apply this method to obtain new solutions, we need to express  the above theorem in terms of the metric (\ref{metricansatze}). By considering the relations (\ref{relationsmetrics})  the above theorem becomes as follows.
 \begin{mytheorem}\label{theorem}
From any solution of the Einstein or the Einstein-Maxwell theory with the metric of the form (\ref{metricansatze}) labeled by a subscript $E$ and a possible Maxwell field for the Einstein-Maxwell case, i. e., $A_E=A_0(\rho,z) dt+A_3(\rho,z)d\phi$, the corresponding BD or BD-Maxwell solution can be obtained by the following transformations:
 \begin{eqnarray}
&&W=W_E^{k},\quad \phi=W_E^{1-k}, \quad \mathcal{A}=\mathcal{A}_E, \quad A=A_E,
\nonumber\\
&&U=U_E-\frac{1}{2}\ln \phi, \quad 
K=K_E+\frac{2\omega-1-(2\omega +3)k}{4}\ln\phi.
\end{eqnarray}  
\end{mytheorem}    
The only possible technical difficulty  in using the above method is to put the GR solutions into the particular form (\ref{metricansatze}), since the stationary axisymmetric solutions can be expressed in various  coordinate systems.
For convenience, let us present the corresponding line element in the Jordan frame explicitly as
\begin{eqnarray}\label{JBDmetric}
ds^2_{JBD}&=&\phi^{-1} \big[-e^{2U_E}(dt+\mathcal{A}_E d\varphi)^2
+e^{2(K_E-U_E)}\phi^{(1-k)(\omega+3/2)}(dr^2+dz^2)
+e^{-2U_E}W_E^2d\varphi^2 \big].
\end{eqnarray}

 Now we can compare this result with the results present in the literature. Note that the above theorem is known in the literature in a more restricted form, in which in the works \cite{Nayak,Singh} the case $K=K_E$ is considered, which relates the parameter $k$ with the BD parameter, $\omega$ as $k=(2\omega-1)/(2\omega+3)$. 
 The restricted form with $K=K_E$ is  used to construct  Kerr-Newman --type solutions  \cite{KimBH} in BD theory from GR and also Bonnor--type black dihole solution  is discussed in \cite{KimLee}. Note that  a similar construction was used to generate Einstein-Rosen--type gravitational wave solutions in BD theory \cite{DeliceAkyar} for the special case $W=\rho$.

\subsection{Extended TNSR method in the Einstein frame}

It is well known that a conformal transformation of the form
\begin{eqnarray}\label{conftrans}
\tilde g_{\mu\nu}=\phi\, g_{\mu\nu},\quad \psi=\left(\omega+\frac{3}{2} \right)\ln \phi
\end{eqnarray}
brings Jordan frame BD action into the Einstein frame as
\begin{equation}
 S_{EBD}=\int d^4 x \sqrt{-\tilde g} \left(\tilde R-\frac{1}{2} \tilde g^{\mu\nu} \partial_\mu \psi\, \partial_\nu \psi- F_{\mu\nu}F^{\mu\nu}\right).
\end{equation}
Note that we use the unit system where $8\pi\, G=1$. One important property of the above transformation is  that the  Maxwell Lagrangian  is invariant under the conformal transformations in four dimensions. Hence, unlike in the case of other matter fields, the scalar field is not coupled with the Maxwell field and the Einstein frame of BD theory for the Maxwell field is equivalent to Einstein-Maxwell--scalar solutions. 
Now, let us investigate the results of the previous section in the Einstein frame. Recall that for a given axially symmetric solution of  Einstein-Maxwell theory  has a corresponding BD-Maxwell solution given in Theorem (\ref{theorem}). 
After applying  the conformal transformation (\ref{conftrans}) to the metric (\ref{JBDmetric}) and the scalar field, the corresponding solution of BD-Maxwell field equations in the Einstein frame becomes
\begin{eqnarray}
{ds}^2_{EBD}&=&-e^{2U_E}(dt+\mathcal{A}_E d\varphi)^2 
+ e^{2(K_E-U_E)+(1-k)\psi}(d\rho^2+dz^2)
+ e^{-2U_E}\,W_E^2\,d\varphi^2,\label{Eframe}\\
\psi&=&\frac{(1-k)(2\omega+3)}{2}\ln{W_E}.\label{alpha}
\end{eqnarray}  
Clearly, we have the following theorem.
\begin{mytheorem}\label{theorem1}
From any solution of the Einstein or the Einstein-Maxwell theory with the metric of the form (\ref{metricansatze}) labeled by a subscript $E$, and a possible Maxwell field for the Einstein-Maxwell case, i. e., $A_E=A_0(\rho,z) dt+A_3(\rho,z)d\phi$, a corresponding BD or BD-Maxwell solution in the Einstein frame (Einstein-Maxwell--scalar solution) can be obtained by the following transformations: 
 \begin{eqnarray}
&&\tilde W=W_E,\quad \psi=\frac{(1-k)(2\omega+3)}{2}\ln{W_E}, \quad 
\mathcal{\tilde A}=\mathcal{A}_E, \quad \tilde A=A_E,
\\
&&\tilde U=U_E,\quad \tilde K=K_E+\frac{1-k}{2}\psi.\nonumber
\end{eqnarray}  
\end{mytheorem}  
Note that a solution generating technique was presented for Einstein-Maxwell--scalar (EMS) theory in \cite{ErisGurses} where in that work it was shown that, for a given Einstein-Maxwell solution, the corresponding EMS solution can be found by solving a set of differential equations involving only $K$ and $\psi$ in our notation.

\subsection{Examples in fordan Frame}

\subsubsection{Flat spacetime}

Let us apply the above procedure to flat Minkowski spacetime, to demonstrate the method and clarify some issues in the previous restricted form of this technique. If we consider Minkowski spacetime in Cartesian coordinates,  we see that since $W_E=1$, the resulting solution is also the same flatspacetime in Minkowski coordinates, since $\phi=W_E^{1-k}$ becomes constant. In order to obtain a nontrivial scalar field, we have to use coordinate systems where $W_E$ is not a constant, such as  the flat spacetime given in spherical coordinates:
\begin{eqnarray}
ds^2=-dt^2+dr^2+r^2(d\theta^2+\sin^2\theta d\varphi^2).
\end{eqnarray}
Note that here $r$ denotes spherical radial coordinate. In order to bring this metric into the form (\ref{metricansatze}), we can consider the transformation $r=e^R$, which brings the metric into the form as
\begin{eqnarray}
ds^2=-dt^2+e^{2R}(dR^2+d\theta^2+\sin^2\theta d\varphi^2).
\end{eqnarray}
Here, by comparing this with (\ref{metricansatze}), we see that it resembles  the form with the identification $R=\rho$, $\theta=z$, and
\begin{equation}
W_E=e^{R}\sin\theta,\quad U_E=0,\quad K_E=R, \quad \mathcal{A}_E=A_E=0.
\end{equation}
Now applying the procedure in  Theorem \ref{theorem}, we obtain
\begin{eqnarray}\label{sphericalflat}
ds^2&=&(r\sin\theta)^{k-1}\big[-dt^2+(r \sin\theta)^{(1-k)^2(\omega+3/2)} 
(dr^2+r^2 d\theta^2)+ (r\sin\theta)^2 d\phi^2  \big],\\
\phi&=&(r\sin\theta)^{1-k}.
\end{eqnarray}
Clearly, the application of this method breaks the spherical symmetry of the BD empty spacetime, as discussed in \cite{KimLee}.  Namely, even if one starts with a spherically symmetric static solution as a seed, the resulting metric will be also static but axially symmetric. Thus, this method is not appropriate  for obtaining the spherically symmetric BD solutions and one may use the methods given, for example,  in \cite{Janis} to obtain spherically symmetric solutions. The reason is that the $g_{zz}$ and $g_{\varphi \varphi}$ components are changed after the application of this method.

One can also consider the cylindrical flat spacetime  given by
\begin{eqnarray}
ds^2=-dt^2+d\rho^2+dz^2+\rho^2 d\varphi^2,
\end{eqnarray}
as the seed solution with $K_E=U_E=0, W_E=\rho$. After applying the algorithm, the resulting flat space solution of BD theory becomes
\begin{equation}
ds^2=\rho^{k-1}\left[-dt^2+ \rho^{(1-k)^2(\omega+3/2)}\left(d\rho^2+dz^2\right)+\rho^{2}d\varphi^2 \right].
\end{equation}
Note that one can also obtain this solution from (\ref{sphericalflat}) by usual transformation relations between spherical and cylindrical coordinates.  Clearly, unlike spherical symmetry, the cylindrical symmetry is preserved by this method. 

\subsubsection{Bonnor type dipole solution}

As we have discussed, this method cannot be used to construct spherically symmetric solutions even if one starts with a spherical seed solution. The resulting solution becomes axially symmetric. Thus, we cannot obtain a Schwarzschild-type solution using this algorithm. However, for static axially symmetric solutions we can use this method to obtain new solutions with the same symmetry. Using this fact, we now generalize a previous BD solution constructed using the restricted form of the TNSR algorithm to its more general form. The solution we choose as the seed solution is a static  axially symmetric solution of Bonnor \cite{Bonnor} representing a magnetic dipole in GR. 
The corresponding BD solution of the magnetic dipole  presented by Bonnor as a solution of Einstein-Maxwell theory  was presented in \cite{KimLee} using the TNSR algorithm. Since the original method is  restricted,  we now present the more general  version of this solution using the extended  version we have discussed.
The GR solution reads
\begin{eqnarray}
ds^2&=&\left(1-\frac{2 M r}{\Sigma} \right)^2\bigg [-dt^2 
+\frac{\Sigma^4}{\left[\Delta+\left(M^2+a^2 \right)\sin^2\theta \right]^3} \left(\frac{dr^2}{\Delta}+d\theta^2 \right) \bigg]
+ \frac{\Delta \sin^2\theta}{\left( 1-\frac{2Mr}{\Sigma}\right)^2} d\varphi^2, \\
 A_0&=&0,\quad  A_3=\frac{2 a M r \sin^2\theta}{\Delta+a^2\sin^2\theta},\quad 
\Delta=r^2-2M r+a^2, ~~~~\Sigma=r^2-a^2\cos^2\theta. 
\end{eqnarray}
In order to apply the algorithm,  we need to bring the metric into the form of (\ref{metricansatze}). To do this we need to apply a transformation $r=e^R+M+(M^2+a^2)e^{-R}$, bringing the $(r, \theta)$ sector of the metric into the form $dr^2/\Delta+d\theta^2=dR^2+d\theta^2.$ Hence by comparing the obtained metric with (\ref{metricansatze}), we can read the metric functions of Einstein-Maxwell theory as follows:
\begin{eqnarray}
&&e^{2U_E}=\left(1-\frac{2mr}{\Sigma} \right)^2,~~ \mathcal{A}_E=0,~~W_E^2=\Delta \sin^2\theta\nonumber\\
&& e^{2K_E}=\frac{\Sigma^4}{\left[\Delta+\left(M^2+a^2 \right)\sin^2\theta \right]^3}\left(1-\frac{2mr}{\Sigma} \right)^4,
\end{eqnarray}
By applying the extended TNSR algorithm and transformation back to the Boyer-Lindquist coordinates, the BD version of the Bonnor's solution becomes
\begin{eqnarray}
ds^2&=&\left(\sqrt{\Delta}\sin\theta \right)^{k-1} \bigg\{ \left(1-\frac{2 M r}{\Sigma} \right)^2\bigg [-dt^2 
+\frac{\Sigma^4\, \left(\sqrt{\Delta}\sin\theta \right)^{(1-k)^2(\omega+3/2)}   }{\left[\Delta+\left(M^2+a^2 \right)\sin^2\theta \right]^3} \left(\frac{dr^2}{\Delta}+d\theta^2 \right) \bigg]
  +\frac{\Delta \sin^2\theta}{\left( 1-\frac{2Mr}{\Sigma}\right)^2} d\varphi^2\bigg\},\quad \ \ \\
 \phi&=&\left(\sqrt{\Delta}\sin\theta \right)^{1-k},\quad A=A_3 d\varphi.
\end{eqnarray}
Note that this solution reduces to one given in \cite{KimLee} for the restricted case given by  $k=(2\omega-1)/(2\omega+3)$.

\section{A  General Axially Symmetric Solution: Brans-Dicke-Kinnersley solution}
\subsection{Brans-Dicke-Kinnersley solution}
The general Einstein-Maxwell type $D$ solution, known in the literature as the Plebanski-Demianski solution \cite{Plebansky}, first presented by Debever \cite{Debever} is given by the metric \cite{Stephani}
\begin{eqnarray}
ds^2&=&(1-pq)^{-2}\bigg[\frac{X\left(dt+q^2 d\sigma \right)^2-Y\left(dt-p^2d\sigma \right)^2}{p^2+q^2} 
+\left(p^2+q^2 \right)\left( \frac{dp^2}{X}+\frac{dq^2}{Y}\right) \bigg],\\
X&=&\left(-g^2+\gamma-\Lambda/6\right)+2 l p-\epsilon p^2+2 m p^3 
-\left(e^2+\gamma+\Lambda/6 \right)p^4,\\
 Y&=&\left(e^2+\gamma-\Lambda/6\right)-2 m q+\epsilon q^2-2 l q^3 
+\left(g^2-\gamma+\Lambda/6 \right)q^4.
\end{eqnarray}
Here $m$ is mass, $l$ is the Newman-Unti-Tamburino (NUT) parameter, $\gamma$ and $\epsilon$ are related to the angular momentum per unit mass $a$,  and the acceleration $b$ and $e$ and $g$ are the electrical and magnetic charges.  
Now we want to find the BD version of this solution.  Thus, we need to cast the metric to  the form given in (\ref{metricansatze}). To do this, since the method do not work in the presence of a cosmological constant, we set $\Lambda=0$ \cite{Kinnersley}. We  also factorize the $dt, d\sigma$ sector of the metric and apply the following coordinate transformations $dp=\sqrt{X}d\theta$, $dq=\sqrt{Y}dr$, $\sigma=\varphi$, then the metric components of GR for this solution in the form (\ref{metricansatze}) becomes
\begin{eqnarray}
&& 2\, K_E=\ln\left[\frac{Y-X}{(1-pq)^4}\right],\quad ~~~~~~~~~~ A_E=\frac{q^2X+p^2Y}{X-Y}
 ,\nonumber \\ 
&&2\, U_E=\ln\left[\frac{Y-X}{(1-pq)^2\left(p^2+q^2 \right)} \right],~
W_E=\frac{\sqrt{XY}}{(1-pq)^2} .
\end{eqnarray}
By applying the method given in Theorem \ref{theorem}, and transforming back to the $p,q$ coordinates, the BD version of the Plebanski-Demianski solution with a vanishing cosmological constant is found to be of the form
\begin{eqnarray}
ds^2&=&\left[\frac{\sqrt{XY}}{\left(1-pq\right)^{2\left(1+\frac{1}{k-1}\right)}}\right]^{k-1}
\left\{   \frac{X\left(dt+q^2 d\sigma \right)^2-Y\left(dt-p^2d\sigma \right)^2}{p^2+q^2} \right. \\ &&
\left. +\left(p^2+q^2 \right)\left( \frac{dp^2}{X}+\frac{dq^2}{Y}\right) \left[\frac{\sqrt{XY}}{\left(1-pq\right)^2}\right]^{(1-k)^2\left(\omega+\frac{3}{2}\right)}  \right\},~~~  
\phi=\left[\frac{\sqrt{XY}}{\left(1-pq\right)^2}\right]^{1-k}.
\end{eqnarray}
As to the best of our knowledge, this solution for BD theory is new. Note that, setting the acceleration parameter to zero and employing Boyer-Lindquist--type coordinates, one can recover a BD Kerr-Newman-NUT--type solution.  Note  alsothat, due to Hawking's theorem \cite{Hawking}, these solutions do not describe asymptotically flat black holes originated from the gravitational collapse of stars, unless the scalar field is a constant. However, these solutions may represent the exterior fields of compact bodies whose interior is characterized by the  parameters of this solution. Hence, it is still important to investigate certain properties of this solution.

\subsection{Kerr-Newman-Taub-NUT solution in BD theory \label{KNTN}}

Here we obtain the BD version of general Kerr-Newman-Taub-NUT (KNTN) solution using the extended  method. Note that the solution can also be obtained from the Plebanski-Damianski--type solution discussed above by a relevant limiting procedure. In GR, this solution has the metric and Maxwell field as follows:
\begin{eqnarray}\label{KNNUT}
ds^2&=&-\left(\frac{\Delta-a^2\sin^2\theta}{\Sigma}\right)\left(dt+\mathcal{A} d\phi \right)^2 
+\frac{\Delta\sin^2\theta}{\Sigma}\left[\frac{(r^2+a^2+n^2 -a b)^2}{\Delta-a^2\sin^2\theta}d\phi^2 \right]
+\Sigma\left( \frac{dr^2}{\Delta} + d\theta^2 \right),~~~~\\
A_E&=&-\frac{q\, r}{\Sigma}(dt-b\, d\phi),
\end{eqnarray}
where the metric functions are given by
\begin{eqnarray}
\Delta&=& r^2-2m\,r+a^2-n^2+q^2,\quad 
\Sigma=r^2+(n+a \cos\theta)^2,\\
\mathcal{A}&=& \frac{a(r^2+a^2+n^2)\sin^2\theta-b\,\Delta}{\Delta-a^2\sin^2\theta},\quad 
b= a \sin^2\theta-2 n \cos\theta. \nonumber
\end{eqnarray}
 By transforming the radial coordinate as $r=e^R+m+(m^2-a^2+n^2-q^2)/4\ e^{-R}$ and using the transformation given in \cite{Misra}, the  $(r,\theta)$,  sector of the metric becomes $dr^2/\Delta+d\theta^2=dR^2+d\theta^2$ such that it resembles the form of the metric (\ref{metricansatze}) with the identification $R=r$, $\theta=z$. Then, one can apply the above theorem and find the corresponding solutions in BD theory easily in these coordinates. Transforming back to the original coordinates, the KNTN solution in BD theory can be expressed in Boyer-Lindquist--type coordinates as follows:
 \begin{widetext}
 \begin{eqnarray}
ds^2&=&(\sqrt{ \Delta} \sin\theta)^{k-1} \Bigg\{-  \left(\frac{\Delta-a^2 \sin^2\theta}{\Sigma}\right)\left(dt+\mathcal{A} \,d\phi \right)^2
+\frac{\Delta\sin^2\theta}{\Sigma}\left[\frac{(r^2+a^2+n^2 -a b)^2}{\Delta-a^2\sin^2\theta}d\phi^2 \right] \nonumber\\ 
&+& \left( \sqrt{\Delta} \sin\theta \right)^{(1-k)^2(\omega+\frac{3}{2})} \Sigma\left( \frac{dr^2}{\Delta} + d\theta^2 \right) \Bigg\},\\
A&=&A_E.
\end{eqnarray}
\end{widetext}
This solution is also presented in \cite{Park} by using a sigma-model analysis with having  a different constant parameter $\alpha$ where the relations of this parameter and the parameter $k$ are given by $\alpha=(1-k)(2\omega+3)/4$.   
The  Ricci scalar of this solution is given by
\begin{eqnarray}
R&=&(k-1)^2\omega \frac{\Delta \cos^2\theta+(M-r)^2\sin^2\theta}{
\Sigma}
\left(\sqrt{\Delta}\sin\theta \right)^{k-3-\frac{1}{2}(k-1)[1-2\omega+k(2\omega+3)]},
\end{eqnarray}
which is helpful to better understand spacetime singularity structure and it 	vanishes  for the limit $k\rightarrow 1$. Clearly, there is a ringlike singularity at the location described by $\Sigma=0.$ The last term in the parentheses vanishes if its exponent becomes zero.
Note that  the parameter range of this spacetime is investigated \cite{Park} for the existence of a black hole type solution with a regular horizon by considering some regularity and existence conditions and concluded that for $\omega<-3/2$, $\alpha_\pm=(1\pm \sqrt{1+4|2\omega +3|})/4$  spacetime may represent a black hole with a regular horizon. In our notation the latter  condition is $k_\pm=1-(1+\sqrt{1+4|2\omega+3|})/(2\omega+3) $.  Note that the symmetry axis is non-null for $\alpha_-$ but null for $\alpha_+$. In these black hole--type solutions, there is a caveat: the scalar field do not satisfy the weak energy condition. Hence, these black holes do not satisfy Hawking's criteria \cite{Hawking}. Moreover, this black hole has a vanishing surface gravity and they are ergocold since they have zero Hawking temperature \cite{KimBH,Park}.

	\subsection{Circular geodesics of the BD-KNTN Spacetime}
	
	Most of the previous studies of axially symmetric spacetimes in BD theory are focused on the obtaining the solutions and analyzing the solutions for whether these solutions represent black holes. Another  property of these solutions  with possible astrophysical importance is the  motion of the  test particles.  There are few works on this issue \cite{Dereli}. In this paper we have also investigated the circular geodesics of  BD-KNTN solution. However, since the calculations and equations are rather complicated and lengthy, we present them in the Appendix. Here below we summarize our general resuls.

  We have first calculated the general geodesics equations in the case of equatorial motion. Then we have considered null and timelike geodesics. For photons, we have  reduced general radial geodesics equations  to  integrals depending on the radial coordinate in the special case $L=aE$  where $E$ is the energy and $L$ is the angular momentum of the test particles or photons. For general case, we have been able to obtain the equation of the photon sphere and realized that it is the same as the corresponding GR case. For timelike particles, we have obtained an effective potential equation for radial motion. We have analyzed this equation for the existence of the inner most stable circular orbit (ISCO). After long and complicated calculations, we are able to obtain the equation determining ISCO. This equation has been analyzed for the special cases such as the extremal case and the BD-Kerr case, as well. These equations agree with the corresponding limiting cases in GR.

\section{Magnetized Solutions in Brans-Dicke Theory}

Obtaining a magnetized solution from a given  seed  solution using Ehlers-Harrison--type transformations  has a long history \cite{Ernstmag}. Using these transformations one can generate a magnetized solution from a given Einstein-Maxwell solution.  For example for a given axisymmetric solution with property $g_{i\varphi}=0$, the following transformations \cite{Dowker} bring a given solution of EMS solutions (i.e., the BD-Maxwell vacuum solutions in the Einstein frame) into the  magnetized one:
\begin{eqnarray*}
&& g'_{ij}=\Lambda^2 g_{ij}\quad (i,j\neq \varphi),\quad  g'_{\varphi\varphi}=\Lambda^{-2} g_{\varphi\varphi},  \alpha'=\alpha, 
\\
&& A'_{\varphi}=-\frac{2}{\Lambda B} \left(1+\frac{1}{2}B A_\varphi \right)+ \frac{2}{B}, \quad 
\Lambda=\left(1+\frac{1}{2}B A_{\varphi} \right)^2+\frac{1}{4} B^2 \, g_{\varphi\varphi}.
\end{eqnarray*} 
 One can also transform the obtained solutions  to the Jordan frame by using (\ref{conftrans}) to obtain magnetized BD solutions as well.  This method is used for BD theory \cite{KimLee} in  connection with the previous version of the TNSR method \cite{Nayak,Singh} to obtain a magnetized black dihole--type solution. 
However, if a  magnetized solution is already presented in the Einstein-Maxwell theory, we can use the extended TNSR method  to obtain the corresponding magnetized solution in BD theory for Jordan or Einstein frames considering Theorem  \ref{theorem} or \ref{theorem1}.  

Using the extended TNSR method discussed above, we can  now present some new magnetized BD solutions such as the magnetized  Kerr-Newman--type solution or the magnetized Bonnor dihole solution. 

\subsection{Magnetized Kerr-Newman solution}

The magnetized Kerr-Newman solution is first discussed in \cite{Ernstmag,Ernstmag1} but since the  Harrison transformations for stationary metrics are more complicated and cumbersome then the  static ones, the full exact solution is first presented recently by \cite{GibbonsMujataPope}. In their notation, the solution reads
\begin{eqnarray}
ds^2&=&H\left[ -f dt^2+R^2\left(\frac{dr^2}{\Delta}+d\theta^2\right) \right] 
+ \frac{\Sigma \sin^2 \theta}{H R^2} \left(d\varphi-\Omega dt \right)^2,\\
A_E&=&\Phi_0+ \Phi_3 \left(d\varphi-  \Omega dt \right) 
\end{eqnarray}
where
\begin{eqnarray}
&&R^2 =r^2+a^2\cos^2\theta\,,~~ 
        \Delta= (r^2+a^2) - 2m r + q^2+p^2,\nonumber\\
&&  f= \frac{R^2 \Delta}{\Sigma}\,,\qquad
~~~~~~~~\Sigma=(r^2+a^2)^2 - a^2\Delta\sin^2\theta,\label{knfns}
\end{eqnarray}
and the (very lengthy) expressions of the other metric and field variables $H,\Omega(\equiv \omega$ in their paper), $\Phi_0,\Phi_3$ can be found in \cite{GibbonsMujataPope}. 

To use the extended TNSR method, we need to bring the above metric into (\ref{metricansatze}), and  to do this we should regroup the $t, \varphi$  sector of the metric and consider the transformation $dr/\sqrt{\Delta}=dR$, then we can read off the metric functions of the GR solution as
\begin{eqnarray}
&& e^{2U_E}=Hf-\frac{\Sigma \Omega^2\sin^2\theta }{H R^2} , \qquad e^{2K_E}=H R^2 e^{2U_E}, \quad 
\mathcal{A}_E=\frac{\Omega \Sigma \sin^2\theta }{f H^2 R^2-\Sigma \Omega^2\sin^2\theta},\quad ~ W_E= \sqrt{\Delta} \sin{\theta} . 
\end{eqnarray}
Then applying the extended TNSR method, the corresponding magnetized Kerr-Newman type solution for BD theory in the Jordan frame can be obtained as
\begin{widetext}
\begin{eqnarray}
ds^2&=&\left(\frac{\sqrt{f\Sigma} \sin\theta}{R}\right)^{k-1}
\Bigg\{
H\bigg[ -f dt^2+R^2 \left(\frac{\sqrt{f\Sigma} \sin\theta}{R}\right)^{(k-1)^2\left(\omega+\frac{3}{2} \right)} \left(\frac{dr^2}{\Delta}+d\theta^2\right) \bigg] + \frac{\Sigma \sin^2 \theta}{H R^2} \left(d\varphi-\Omega dt \right)^2
\Bigg\}.\ \ \\
\phi&=&\left(\frac{\sqrt{f\Sigma} \sin\theta}{R}\right)^{1-k}
\end{eqnarray}
\end{widetext}
Note that the Einstein frame solution can be obtained by  considering only the terms in the curly bracket and replacing $\phi$ with $\psi$ as in (\ref{Eframe}) and (\ref{alpha}). The physical properties of this solution will be treated elsewhere.

\subsection{Magnetized Bonnor dihole solution}

Now as a final application of the method discussed in this paper, we present a two parameter extended version of the BD solution \cite{KimLee}  for   Bonnor-type dihole solution \cite{Bonnor} embedded in a Melvin magnetic universe presented in \cite{Emparan}.
Starting from Einstein-Maxwell magnetized solution and applying the Theorem \ref{theorem1}, we obtain the following solution:
\begin{widetext}
\begin{eqnarray}
ds^2&=&\left(\sqrt{\Delta}\sin\theta \right)^{k-1} \left\{ \Lambda^2\left[-dt^2
+\frac{\Sigma^4\, \left(\sqrt{\Delta}\sin\theta \right)^{(1-k)^2(\omega+3/2)}   }{\left[\Delta+\left(M^2+a^2 \right)\sin^2\theta \right]^3} \left(\frac{dr^2}{\Delta}+d\theta^2 \right) \right]+ \frac{\Delta \sin^2\theta}{\Lambda^2} d\varphi^2\right\},\quad \ \ \\
\phi&=&\left(\sqrt{\Delta}\sin\theta \right)^{1-k},\\
 A&=&A_3d\varphi=-\frac{2Mr a+B/2\left[\left(r^2-a^2 \right)^2+\Delta a^2 \sin^2\theta\right]}{\Lambda\, \Sigma}\sin^2\theta\, d\varphi,\\
\Lambda&=&\frac{\Delta+a^2\sin^2\theta+2B M r a \sin^2\theta+B^2/4 \sin^2\theta\left[\left(r^2-a^2 \right)^2+\Delta a^2 \sin^2\theta\right]}{\Sigma}.
\end{eqnarray}
\end{widetext}

\section{General Relativistic Limit}

The GR limit of the solutions of the BD theory is not as trivial as it sounds. It was often claimed that as $\omega\mapsto \infty$ BD solutions reduce to the corresponding GR solutions. However,  several counterexamples of this claim are present in the literature. For example, for the special case of vanishing trace of the energy-momentum tensor, $T$, several papers \cite{romero-barros,banerjee-sen,faraoni-w-dependence,bhadra-nandi,dilek,Baykal} are showed that this is not true in general. This fact is related  in \cite{faraoni-w-dependence} with  
 the conformal invariance of BD theory  for vanishing $T$. This may be also due to fact that the scalar field equation becomes sourceless and the integration constant of this scalar equation becomes independent of $\omega$ since the remaining  field equations cannot fix this constant in terms of $\omega$. A perfect example is a recent paper \cite{dilek}, in which BD-Maxwell solutions for higher dimensional static  cylindrically symmetric spacetime were obtained. In these solutions, when the spacetime dimensions is greater than 4, the trace $T$ does not vanish and this yields an equation relating BD parameter $\omega$ and the  extra integration constant  of the solution and solving this constant for $\omega$, one recovers the corresponding GR solutions \cite{pinar} in the $\omega \mapsto \infty$ limit. However, in four dimensions, the field equations do not yield such an equation and  for this case \cite{Baykal} the GR limit is achieved not by  $\omega\mapsto \infty$ but a specific value of this integration constant where the BD scalar becomes a constant.  Even for a nonvanishing $T$, some counterexamples were found \cite{Chauvineau} 
where the scalar field does not go to a constant sufficiently fast and therefore the  GR limit is not recovered as $\omega \mapsto \infty$  for these solutions.

 For the cases where this limit do not arise as a result of field equations,  some physical arguments  can be put forward to fix this extra constant in terms of $\omega$. However, these arguments  are not in general  the result of encompassing physical  requirements  but the results of some special considerations. For example in \cite{bhadra-nandi}  where some spherically symmetric BD solutions were discussed, the extra constant is fixed by demanding the matching of the solution to an internal source in the weak field approximation. Although the  GR limit  is obtained as $\omega\mapsto \infty$ for this special case, one cannot ensure that this procedure also works for the full theory, especially in the presence of  a source having a strong field.
 Thus, in order to investigate the GR limit of BD solutions, not only  mathematical results but also additional physical arguments  might be required, as in \cite{bhadra-nandi}.  
 
 In order to avoid all these complexities in this paper, when we have discussed the GR limit, we have only considered the case where the scalar field becomes a constant. As it is clear from Eq. (\ref{trans2}), as $k\rightarrow 1 $, the stationary axially symmetric BD(-Maxwell) vacuum  solutions reduce to GR for \emph{any finite}  $\omega$ irrespective of  whether it is large or small, since $\phi$ becomes a constant. Note that, similar to the case of generic finite values of $\omega$ where the BD theory is equivalent to GR with a minimally coupled scalar field, modulo potential singularities in transformation between frames, when $\omega\mapsto \infty$,  it is known that the theory reduces to GR plus a minimally coupled (residual) scalar field \cite{chauvineau1} even  if $\lim_{\omega\mapsto \infty} \phi=\phi_0=\mbox{constant}$. Hence, we need some consistency requirements  on the $k$, since this requires $\lim_{\omega \mapsto \infty} k=1$.  For example, the regularity of the spacetime 
 can be ensured by choosing $k$ such that  $\lim_{\omega \mapsto \infty} k \approx 1 - \sqrt{\frac{\beta}{2\omega+3}}$ with $\beta$ an arbitrary constant. In the limit $\omega\mapsto \infty$, for this value of $k$, the metric function $K$ does not approach to its corresponding GR value but becomes $K=K_E+\frac{\beta}{4} \ln W_E $.  Hence, as $\omega \mapsto \infty$, BD solutions discussed in this paper do not approach their corresponding GR solutions. Note also that although for $k \mapsto 1$ the GR limit is obtained since the BD scalar becomes a constant, the additional scalar degrees of freedom is always present  at the level of the  fluctuations, even for the constant background solutions.

\section{Conclusions}
In this paper, we have investigated stationary axially symmetric vacuum  solutions of the  Brans-Dicke-Maxwell theory in both Jordan and Einstein frames. We first show  that, employing a particular form of the standard metric simplifies the procedure of obtaining the Ernst equation from the BD field equations. By investigating the Ernst equation and also the remaining field equations, we are able to obtain a two parameter extension of a particular solution generating technique for BD theory. In order to show how this method works, we have  constructed several known solutions and also some  new solutions for BD theory, such as a Plebansky-Demiansky--type solution with a vanishing cosmological constant  or magnetized Kerr-Newman--type solutions. We have also discussed the GR limit of these solutions in some detail.

  As a further physical application, we have also investigated the geodesics equations for Kerr-Newman-NUT type solution of BD theory in Appendix \ref{circgeo}, where we have focused on the circular null and timelike geodesics of this solution and successfully obtained the ISCO for this spacetime. A question arises whether this known solution generating technique generalized in this paper to two parameters can be applicable to other theories such as f(R)-type theories or EMS-type theories. Especially, EMS-type theories are quite interesting since a regular  black hole type solution with  scalar hair \cite{Boronnikov,bekenstein1,bekenstein2}  exists and interest has recently been focused on the possibility of the existence of stationary-type hairy black hole solutions \cite{Anabalon1,Anabalon2,Astorino1,Maeda,Bardoux,Astorino2}.

\section*{Acknowledgements}
The authors would like to thank Ahmet Baykal and Dilek K. \c{C}iftci for reading the manuscript and for the useful comments. P. K. and O.D. are supported by Marmara University Scientific Research Foundation (Project No. FEN-C-DRP-090414-0094).

\appendix

	\section{CIRCULAR GEODESICS OF THE BD-KNTN SPACETIME
	 \label{circgeo}}
	In order to determine the geodesic equations of BD-KNTN solution given  in Sec. \ref{KNTN}, we first define the Lagrangian of the system as
	\begin{eqnarray}
		\mathcal{L}=\frac{1}{2}\left( g_{tt}\dot{t}^2+g_{t\varphi}\dot{t}\dot{\varphi}+g_{\varphi\varphi}\dot{\varphi}^2+g_{rr}\dot{r}^2+g_{\theta \theta}\dot{\theta}^2 \right),
	\end{eqnarray}
	where  here the overdot  denotes the differentiation with respect to the  proper time $\tau$ for timelike case and an affine parameter for the null case. The symmetries of this Lagrangian enable us to determine  the following first integrals of the motion 
	\begin{eqnarray}\label{EL}
		g_{tt}\dot{t}+g_{\varphi t}\dot{\varphi}=-E,~~~~~~~~g_{t\varphi}\dot{t}+g_{\varphi \varphi}\dot{\varphi}=L,
	\end{eqnarray}
	where the constants $E$ and $L$ are related with energy and angular momentum of the particle.	 Furthermore, using the metric itself,  we have
	\begin{eqnarray}\label{geodesic}
		\epsilon=g_{tt}\dot{t}^2+2g_{t\varphi}\dot{t}\dot{\varphi}+g_{\varphi \varphi}\dot{\varphi}^2+g_{rr}\dot{r}^2+g_{\theta \theta}\dot{\theta}^2,
	\end{eqnarray}
	where $\epsilon=1,0,-1$ for spacelike, lightlike and timelike geodesics, respectively.
	From (\ref{EL})  we obtain
	\begin{subequations}
		\begin{align}
		&	\dot{t}=\frac{-e^{2U}LB-e^{2U}EB^2+e^{-2U}EW^2}{W^2},\\
		&	\dot{\varphi}=\frac{e^{2U}(L+EB)}{W^2}.
		\end{align}
	\end{subequations}
	To determine the geodesics in the equatorial plane, we set $\theta=\frac{\pi}{2}$ and $\dot{\theta}=0$, and  also use (\ref{KNNUT}), then the above equations become
	\begin{subequations}
		\begin{align}
		&	\dot{\varphi}=\frac{(\Delta-a^2)(L-a E)+aE(r^2+n^2)}{(r^2+n^2) \Delta^{(k+1)/2}},\label{kntngp}\\
		&	\dot{t}=\frac{1}{(r^2+n^2) \Delta^{(k+1)/2}} \left[a(\Delta-a^2)(L-a E )
			+(r^2+n^2)(E(r^2+n^2)-a(L-2E a))\right]\label{kntngt}.
		\end{align}
	\end{subequations}
	Using these expressions in (\ref{geodesic}), we obtain the radial equation determining the geodesics of BD-KNTN spacetime as
	\begin{eqnarray}
		\dot{r}^2&=&\frac{\Delta e^{-2K}\left[-e^{4U}(L+EB)^2+E^2W^2+\epsilon\, W^2e^{2U}\right]}{W^2}\nonumber\\
		&=&\frac{(a^2-\Delta)(L-a E)^2+(r^2+n^2)\big[E^2 (r^2+n^2)+2aE(a E-L)+\epsilon  \Delta^{(k+1)/2}\big]}{(r^2+n^2)^2\Delta^{\frac{1-k}{4}[2\omega-1-k(2\omega+3)]}}.\label{kntngr}.
	\end{eqnarray}
	These last three equations are the general geodesics equations for equatorial motion. Now, let us consider null and timelike motion in some detail. These equations reduce to corresponding GR  ones \cite{Chakraborty,Pradhan} for $k\mapsto 1$.
	
	\subsection{Light-like geodesics ($\epsilon=0$)}
	Null geodesics are obtained by setting the parameter $\epsilon=0$, then the radial equation becomes
	\begin{eqnarray}\label{kntngr1}
\dot{r}^2&=&\frac{\Delta e^{-2K}\left[-e^{4U}(L+EB)^2+E^2W^2+\epsilon\, W^2e^{2U}\right]}{W^2}\nonumber\\
		&=&\frac{(a^2-\Delta)(L-a E)^2+(r^2+n^2)\big[E^2 (r^2+n^2)+2aE(a E-L)\big]}{(r^2+n^2)^2\Delta^{\frac{1-k}{4}[2\omega-1-k(2\omega+3)]}}.		
	\end{eqnarray}
	In order  to distinguish the geodesics with different impact parameter $D=L/E$, we consider  the following cases:
	\begin{itemize}
		\item[(a)]
		The special case $L=a E$:
		
		For this case the impact parameter becomes $D=a$, which plays an important role to study the radial geodesics. Equations (\ref{kntngp}),(\ref{kntngt} and (\ref{kntngr}) become
		\begin{subequations}
			\begin{align}
			&	\dot{\varphi}=\frac{a E}{\Delta^{(k+1)/2}}\label{nulvarphi}\\
			&	\dot{t}=\frac{(a^2+r^2+n^2)E}{\Delta^{(k+1)/2}}\label{nult}\\
			&	\dot{r}=\pm E \Delta^{\frac{k-1}{8}(2\omega-1-k(2\omega+3))}\label{kntnglr}
			\end{align}
		\end{subequations}
		Here the positive  sign represents outgoing photons whereas the negative sign represents ingoing photons. Using these equations we find that 
		\begin{subequations}
			\begin{align}
			&	\pm\,  \varphi =a \int   \Delta^{\frac{1}{8}\left[(k-1)^2(2\omega+3)-8\right]}\, dr,\\
			&	\pm\, t	=\int (a^2+r^2+n^2) \Delta^{\frac{1}{8}\left[(k-1)^2(2\omega+3)-8\right]}\, dr
				,\\
			&	\pm\,	 E\, \tau=\int \Delta^{\frac{1-k}{8}[2\omega-1-k(2\omega+3)] }dr.
			\end{align}
		\end{subequations}
		These equations are more complicated than similar equations for GR case \cite{Chakraborty,Pradhan}. For example, unlike GR case, the radial coordinate is not changing uniformly in BD theory.		
		
		\item[(b)] The general case $D=L/E$:
		
		Using Eq. (\ref{kntngr1}), we obtain  the equations corresponding the radius $r_c$ of the circular photon orbit at $E=E_c$ and $L=L_c$  as
		\begin{eqnarray}\label{kntngrt}
			&&(r_c^2+n^2)\dot{r}^2=\frac{1}{\Delta^{\frac{k-1}{4}[2\omega-1-k(2\omega+3)]}}\times\bigg[(r_c^2+n^2)E_c^2\nonumber\\
			&&+\frac{(a^2-\Delta)(aE_c-L_c)^2}{(r_c^2+n^2)}-2aE_c(L_c-aE_c)\bigg]=0.
		\end{eqnarray}
		For the regions where $\Delta\neq 0$, the derivative of the above equation becomes
		\begin{eqnarray}
			&&\Delta^{\frac{1-k}{4}[2\omega-1-k(2\omega+3)]}\times\bigg\{ r_c\,E^2-\frac{(L_c-a\,E_c)^2}{r_c^2+n^2}
			\left[(r_c-m)(r_c^2+n^2)+r_c(a^2-\Delta)\right]\bigg\}=0.~~~~~
		\end{eqnarray}
		Substituting $D_c=\frac{L_c}{E_c}$ into the above equations we obtain
		\begin{eqnarray}
			&&(r_c^2+n^2)+\frac{(a^2-\Delta)(a-D_c)^2}{(r_c^2+n^2)}
			-2a(D_c-a)=0, ~ \label{imp1}\\
			&&r_c-\frac{(D_c-a)^2}{(r_c^2+n^2)^2}\times \big[(r_c-m)(r_c^2+n^2) 
			+r_c(a^2-\Delta)\big]=0\label{imp2}.~~
		\end{eqnarray}
		Note that these equations are independent of the BD parameters $k$ and $\omega$, meaning that the BD scalar field does not affect the geodesics of the photons.  The impact parameter can be obtained from Eq. (\ref{imp2}) as
		\begin{eqnarray}
			D_c=a\pm\sqrt{\frac{r_c(r_c^2+n^2)^2}{(r_c-m)(r_c^2+n^2)+r_c(a^2-\Delta)}}.
		\end{eqnarray}
	\end{itemize}
	Inserting this equation into (\ref{imp1}) yields the equation of photon sphere \cite{Chandrasekhar,Vibhadra}  for BD-KNTN spacetime, which turn out to be the same with the corresponding GR solution \cite{Chakraborty,Pradhan}.
	
	\subsection{Timelike geodesics ($\epsilon=-1$) }
	For timelike geodesics, Eqs. (\ref{kntngp}) and (\ref{kntngt}) remain unchanged, but Eq. (\ref{kntngr}) becomes
	\begin{eqnarray}
		(r^2+n^2)\dot{r}^2&=&\frac{1}{\Delta^{\frac{k-1}{4}(2\omega-1-k(2\omega+3))}}\bigg[(r^2+n^2)E^2\nonumber\\
		&&+\frac{(a^2-\Delta)(aE-L)^2}{(r^2+n^2)}-2aE(L-aE)-\Delta^{\frac{k+1}{2}}\bigg]=0.\quad \ \label{kntntlr1}~~~~
	\end{eqnarray}
	Now, as in the null geodesics, let us consider the special case $L=a\,E$ and the general case separately.	
	
	\begin{itemize}
		\item[(a)] Special case, $L=a\,E$:
		
		For this case, the  above equation  becomes
		\begin{eqnarray}
			\Delta^{\frac{k-1}{4}(2\omega-1-k(2\omega+3))}(r^2+n^2)\, \dot{r}^2= (n^2+r^2)E^2-\Delta^{(k+1)/2} ,~~~~
		\end{eqnarray}
		while Eqs.  (\ref{kntngp}) and (\ref{kntngt})   become the same as for the null geodesics (\ref{nulvarphi}) and (\ref{nult}).
		
		\item[(b)] {The general case ($L-a\,E=x$):}
	\end{itemize}
	
	From Eq. (\ref{kntngrt}) with employing a  reciprocal radius function ($r=\frac{1}{u}$) and substituting $L-aE=x$ we find
	\begin{eqnarray}\label{kntngrt1}
		\Delta^{\frac{k-1}{4}[2\omega-1-k(2\omega+3)]}(1+u^2n^2)^2\,u^{-4}\,\dot{u}^2&=&
		 E^2(1+n^2u^2)^2-2aExu^2(1+n^2u^2)\nonumber\\
		\quad &&
		-u^2(1+n^2u^2)\Delta^{(k+1)/2}+(a^2-\Delta)u^4x^2=0,
	\end{eqnarray}
	where $\Delta=\frac{1}{u^2}-\frac{2 m}{u}+a^2+Q^2-n^2$ for KNTN solution. The derivative of the above equation becomes
	\begin{eqnarray}\label{kntngrt2}
		&&2n^2u^2E^2(1+n^2u^2)-2au^2xE(1+2n^2u^2)
		-u^2 \Delta^{(k+1)/2}(1+2n^2u^2)+2(a^2-\Delta)u^4x^2,\nonumber\\
		&&-(mu-1)\left[\frac{k+1}{2}(1+n^2u^2)\Delta^{(k-1)/2}+u^2x^2\right]=0.\nonumber\\
	\end{eqnarray}
 Equations (\ref{kntngrt1}) and (\ref{kntngrt2}) can be combined to give
	\begin{eqnarray}
		(1+n^2u^2)^2E^2&=&u^4x^2(a^2-\Delta)-(mu-1)(1+n^2u^2)
		\bigg[\frac{k+1}{2} \Delta^{(k+1)/2}(1+n^2u^2)- x^2u^2\bigg],\nonumber\\
	\end{eqnarray}
	and consequently
	\begin{eqnarray}\label{Eeq}
		2aE\, u^2x(1+n^2u^2)&=& 2(a^2-\Delta)u^4x^2
		- u^2(1+n^2u^2) \Delta^{(1+k)/2}\nonumber\\
		&-&(mu-1)(1+n^2u^2)
		\left[\frac{k+1}{2}(1+n^2u^2)\Delta^{(k-1)/2}+u^2x^2\right].\nonumber\\
	\end{eqnarray}
	By eliminating $E$ between these equations, we obtain the following 	quadratic equation for $u^2x^2$:
	\begin{eqnarray}\label{quadratic}
		\mathcal{A}\,u^4x^4+\mathcal{B}\,u^2x^2+\mathcal{C}=0,
	\end{eqnarray}
	where we have defined
	\begin{eqnarray}
		\mathcal{A}&=&4\Delta\left\{\left[2\Delta u^2+(mu-1)(1+n^2u^2)\right]^2-4\Delta
		a^2u^4\right\},\nonumber\\
		\\
		\mathcal{B}&=&4 \Delta^{(k+1)/2}(1+u^2n^2)\bigg\{
		\left[2\Delta u^2+(mu-1)(1+n^2u^2)\right]^2 -4\Delta
		a^2u^4\nonumber\\
		&&\quad +k\,(mu-1)(1+n^2u^2)\big[(mu-1)(1+n^2u^2)
		+ 2\Delta u^2\big]\bigg\},\\
		\mathcal{C}&=&\Delta^k(1+n^2u^2)^2
		\left[2\Delta
		u^2+(k+1)(mu-1)(1+n^2u^2)\right]^2.~~~~
	\end{eqnarray}
	The discriminant $\mathcal{D}=\mathcal{B}^2-4\mathcal{A}\mathcal{C}$ of this equation is
	\begin{eqnarray}
		\mathcal{D}&=&64 \Delta^{k+2}u^4a^2(1+n^2u^2)^2\bigg\{4a^2\Delta u^4
		+k^2(mu-1)^2(1+n^2u^2)^2\nonumber\\
		&&-\left[2\Delta u^2+(mu-1)(1+n^2u^2)\right]^2\bigg\}.
	\end{eqnarray}
	The solution of the Eq. (\ref{quadratic}) is
	\begin{eqnarray}
		&&x^2u^2=-\frac{\Delta^{(k-1)/2}(1+n^2u^2)}{2}
		\left[1+\frac{k'(Z_{+}+Z_{-})}{2Z_{+}Z_{-}}\pm\frac{2\sqrt{\Delta}au^2\sqrt{k'^2-Z_{+}Z_{-}}}{Z_{+}Z_{-}}\right],\nonumber\\
	\end{eqnarray}
	where
	\begin{eqnarray}
		k'&=&k(mu-1)(1+n^2u^2),\nonumber\\
		Z_{+}Z_{-}&=&(2\Delta u^2+(mu-1)(1+n^2u^2))^2-4\Delta a^2 u^4,\nonumber\\
		Z_{\pm}&=&2\Delta u^2+(mu-1)(1+n^2u^2)\pm2\sqrt{\Delta}au^2,\nonumber
	\end{eqnarray}
	and from the above equation we conclude that
	\begin{eqnarray}
		x&=&\pm\left\{-\frac{\Delta^{(k-1)/2}(1+n^2u^2)}{2u^2}
		\left[1	+\frac{k'\,(Z_{+}+Z_{-})}{2Z_{+}Z_{-}}\pm\frac{2\sqrt{\Delta}\,au^2\sqrt{k'^2-Z_{+}Z_{-}}}{Z_{+}Z_{-}}\right]\right\}^{1/2}.
	\end{eqnarray}
	When we put the values of $x$ into Eq. (\ref{Eeq}), we  obtain the energy  of circular orbit
	\begin{eqnarray}
		E&=&\frac{1}{\sqrt{2(1+n^2u^2)}}\Bigg\{-\Delta^{\frac{k-1}{2}}(k+1)(mu-1)(1+n^2u^2)\nonumber\\
		&&+\frac{(mu-1)(1+n^2 u^2)-u^2(a^2-\Delta)}{Z_+
			Z_-} 
		\bigg[-Z_+
		Z_-+k'(Z_++Z_-)/2
		\pm 2 au^2\sqrt{\Delta}\sqrt{k'^2-Z_+Z_-}\bigg]\Bigg\}^{1/2}.
	\end{eqnarray}
	
	In order to determine the ISCO of this spacetime, we need the explicit form of the effective potential. From Eq. (\ref{kntntlr1}) we obtain
	\begin{eqnarray}
		\frac{1}{2}\dot{r}^2=\frac{E^2-1}{2 \Delta^{\frac{k-1}{4}[2\omega-1-k(2\omega+3)]}}+V_{eff},
	\end{eqnarray}
	where
	\begin{eqnarray}
		V_{eff}&=&\frac{1}{2 \Delta^{\frac{k-1}{4}[2\,\omega-1-k(2\omega+3)]}}\bigg[1+\frac{(a^2-\Delta)(a E-L)^2}{(r^2+n^2)^2}
		-\frac{2aE(L-aE)+\Delta^{(k+1)/2}}{(r^2+n^2)}\bigg].
	\end{eqnarray}
	When the radial derivative of the effective potential vanishes, the particle stays in the circular orbit. For the ISCO equation, the second  radial derivative of $V_{eff}$   must also vanish, i.e., $\frac{\partial^2 V_{eff}}{\partial r^2}=0$. We consider the region  outside the ``horizon" where $\Delta\neq 0$. Then we have
	
	\begin{eqnarray}
		&&-\Delta^{k/2}(n^2+r^2)\bigg\{-2\Delta^2(n^2-3r^2)+\Delta(k+1)
		\left[n^2+(4 m-3r)r\right](n^2+r^2) 	 +(k^2-1)(m-r)^2
		(n^2+r^2)^2\bigg\}\nonumber\\ 		&& +4a E\, x\,\Delta^{3/2}(n^2+r^2)(n^2-3r^2)
		-2x^2\Delta^{3/2}\left[2a^2(n^2-5r^2)-2\Delta(n^2-5r^2)
		+(n^2+r^2)(n^2+8 m r-7r^2)\right]=0.\quad\ \
	\end{eqnarray}
	When we set $Ex$ and $x^2$ using the expression above, we obtain  the  ISCO equation
	\begin{widetext}
		\begin{eqnarray}
			&&(k+1)(n^2+r^2)^2\left[-\Delta m-r(k-1)(m-r)^2\right]+\Delta\big[-4r^3(a^2-\Delta)-4r^5-4n^2r^3+m(n^2+5r^2)(n^2+r^2)\big]\nonumber\\
			&&\times\bigg\{\left[2\Delta r+(m-r)(n^2+r^2)\right]^2-4\Delta r^2a^2+k(m-r)(n^2+r^2)\left[2\Delta r+(m-r)(n^2+r^2)\right]\nonumber\\
			&&\pm2ar\sqrt{\Delta}\bigg[4a^2r^2\Delta-\left[2\Delta
			r+(m-r)(r^2+n^2)\right]^2+k^2(m-r)^2(n^2+r^2)^2\bigg]^{1/2}\bigg\}=0.
		\end{eqnarray}
	\end{widetext}

	Having found the general expression, now let us consider a few special cases. First, we consider the radius of ISCO for  the ``extremal BD--KNTN spacetime," where the parameters are related by
	\begin{equation}
		a^2=m^2+n^2-q^2,
	\end{equation}
	in which  the "horizons" of BD-KNTN spacetime
	\begin{equation}
		r_{\pm}=m\pm\sqrt{m^2+n^2-q^2-a^2}
	\end{equation}
	coincides with the extremal Kerr  radius $r=m$. 
	Defining the reduced quantities
	\begin{eqnarray}
		r_m&=&\frac{r}{m},~~~~~~	n_m=\frac{n}{m},
		q_m=\frac{q}{m},~~~~~~ a_m=\frac{a}{m}=\sqrt{1+n^2-q^2}, \quad 
		\Delta= m^2(1-r_m)^2,
	\end{eqnarray}
	we can express the extremal ISCO equation for BD-KNTN spacetime as follows:
	\begin{eqnarray}
		&&-(k+1)(r_m-1)^2\left[1+(k-1)r_m \right](n_m^2+r_m^2)^2\nonumber\\
		&&+m^6(r_m-1)^4\left[n_m^4+2n_m^2r_m^2(3-4r_m)+r_m^3(4q_m^2-3r_m)\right]\nonumber\\
		&&\times\bigg((k+1)n_m^4+2n_m^2r_m(2+k-3r_m)
	+r_m\left\{ r_m\left[4q_m^2+r_m(-4+2k+r_m-kr_m)\right]\right\}\nonumber\\
		&&\pm 2 r_m\sqrt{1+n_m^2-q_m^2}\bigg[(k^2-1)n_m^4+2n_m^2r_m\big[-2
		+r_m(k^2+3)\big]+r_m^2\left\{-4q_m^2+r_m\left[4+r_m(k^2-1)\right]\right\}\bigg]^{1/2}\bigg)\nonumber\\
		&&=0.
	\end{eqnarray}
	The solution of the above equation is $r_m=1$ or $r=m$ which
	coincides with the ISCO of  extremal Kerr spacetime. For this special case, the ISCO radius of the BD-KNTN solution does not depend on the electrical and NUT charges or the Brans-Dicke scalar field.

	For another special case, we can calculate the ISCO equation for the BD  Kerr solution  by setting $n=0$ and $q=0$. Then we have 
	\begin{eqnarray}
		&&(k+1)\left[-\Delta m-r(k-1)(m-r)^2\right]+\Delta r\big(-4a^2+4\Delta %
		-4r^2+5mr\big) \nonumber\\&&\times\ \big[(2\Delta+r(m-r))^2-4a^2\Delta
		+k\,r(m-r)\left[2\Delta-r(m-r)\right]\pm
		2a\sqrt{\Delta}\nonumber\\
		&&\times\sqrt{4a^2\Delta-(2\Delta+r(m-r))^2+k^2r^2(m-r)^2}\big]=0,~~~~~~~~
	\end{eqnarray}
	where here $\Delta=r^2-2mr+a^2$. 
	Our results are in accordance with corresponding solutions in GR \cite{Chandrasekhar,Chakraborty,Pradhan}.

\end{document}